\documentclass[runningheads]{llncs}
\usepackage{graphicx}
\usepackage{color, soul}
\usepackage{todonotes}
\usepackage{comment}
\usepackage{verbatim}

\begin{document}
\title{From Cyber-Security Deception To Manipulation and Gratification Through Gamification}

\author{Xavier Bellekens\inst{1}, Gayan Jayasekara\inst{1}, Hanan Hindy\inst{1}, Miroslav Bures\inst{1,3}, \\ David Brosset\inst{2}, Christos Tachtatzis\inst{4} and Robert Atkinson\inst{4} }
\authorrunning{Bellekens et al.}
%
\institute{Abertay University, Division of Cyber-Security, Scotland \and 
Chair of Naval Cyber Defense, Ecole navale, France \and 
Czech Technical University in Prague, Dpt. of Computer Science, Faculty of Electrical Engineering, Czech Republic  \and
University of Strathclyde, Dpt. of Electronic \& Electrical Engineering, Scotland}
\maketitle              
\begin{abstract}
With the ever growing networking capabilities and services offered to users,  attack surfaces have been increasing exponentially, additionally, the intricacy of network architectures has increased the complexity of cyber-defenses, to this end, the use of deception has recently been trending both in academia and industry. Deception enables to create proactive defense systems, luring attackers in order to better defend the systems at hand. Current applications of deception, only rely on static, or low interactive environments. In this paper we present a platform that combines human-computer-interaction, analytics, gamification and deception to lure malicious users into selected traps while piquing their interests. Furthermore we analyse the interactive deceptive aspects of the platform through the addition of a narrative, further engaging malicious users into following a predefined path and deflecting attacks from key network systems.

\end{abstract}

\section{Introduction}
Over the last two decades the field of cyber-security has experienced numerous changes associated with the evolution of neighboring fields, such as networking, mobile communications, and recently the Internet of Things (IoT)~\cite{desolda2015mashing}~\cite{hodo2016threat}. Changes in mindsets have also been witnessed, a couple of years ago the cyber-security industry blamed users for their mistakes often depicted as the number one reason behind security breaches. Nowadays, companies are empowering users, modifying their perception of being the weak link, into being the center-piece of security design~\cite{faily2018designing}. Users are by definition ``in control" and therefore a cyber-security asset. Researchers have focused on the gamification of cyber-security elements, helping users to learn and understand the concepts of attacks and threats, allowing them to become the first line of defense to report anomalies~\cite{zhan2018assessment}. However, over the past years numerous infrastructures have suffered from malicious intent, data breaches, and crypto-ransomeware, clearly demonstrating the technical ``know-how" of hackers and their ability to bypass the security measures in place~\cite{hill2018cryptoknight}~\cite{hindy1806taxonomy}. 

\paragraph{}Researchers concentrated on the gamification, learning and teaching theory of cyber-security to end-users in numerous fields through various techniques and scenarios to raise end-user cyber-situational awareness~\cite{almeshekah2016cyber}\cite{almeshekah2014planning}. While empowering the end-users, researchers overlooked hackers, and the potential of using hacker gamification to benefit cyber-security. 
In this paper, we argue that there is an endemic issue in the understanding of hacking practices leading to vulnerable devices, software and architectures. We therefore propose a gamification platform for hackers. The platform is designed with hacker user-interaction and deception in mind. \newline

The contributions of this paper are threefold:
\begin{itemize}
\item Deceptive Interactive Techniques are proposed and defined. This can be transposed to other systems to increase the interactivity of deception systems.
\item An Interactive Deceptive Based Platform is presented. The platform is able to deploy scenarios to engage and deceive malicious users in different contexts. 
\item Deceptive narrative scenarios based on gamification are presented and evaluated. The narrative enables to engage the malicious user to further explore the deceptive scenarios and safeguards key elements of the network.
\end{itemize}

The remainder of this paper is organised as follows; Section~\ref{sec:InteractiveDeception} discusses techniques to engage malicious users through Gamification, Narrative, Manipulation and Gratification. Section~\ref{sec:PlatformArchitecture} introduces the interactive deception platform and its components, followed by Section~\ref{sec:Scenarios} which discusses 3 different scenarios to evaluate the interactivity and deceptiveness of the platform. Furthermore, Section~\ref{sec:Evaluation} discusses the results obtained through the different scenarios, Section~\ref{sec:RelatedWorks} highlight the related works, Section~\ref{sec:KeyTakeAways} discusses the key takeaways from the platform and its limitations and the paper concludes with Section~\ref{sec:Conclusion}.

\section{Interactive Deception Based Defenses}
\label{sec:InteractiveDeception}

     \begin{figure}[h]
    \centering
    \includegraphics[height=0.5\textheight, keepaspectratio=true] 
        {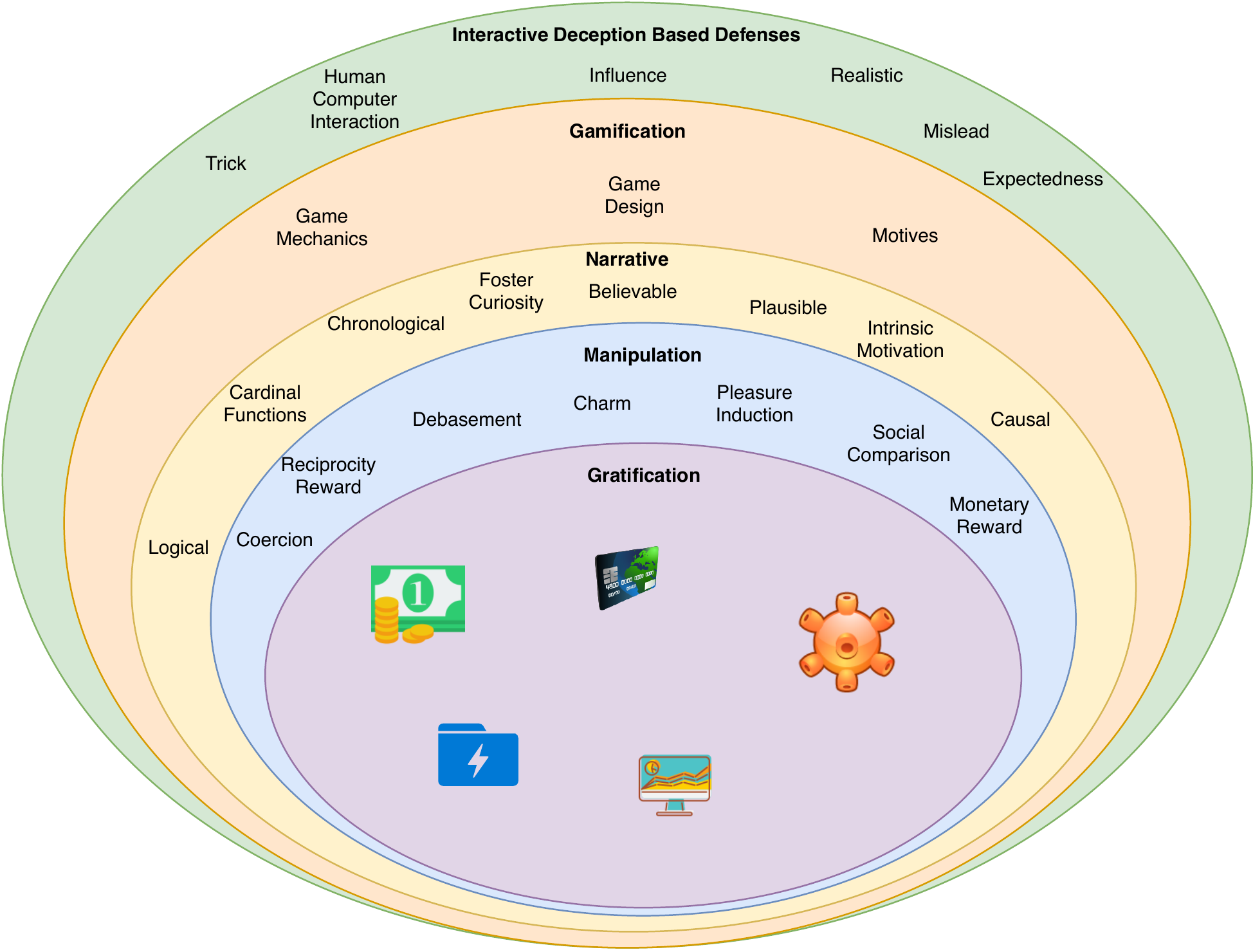}
    \caption{Inner Architecture of the Deception Framework}
      \label{DPA}
    \end{figure}

The premise of this research assumes that all networks are vulnerable. This premise is extended to the belief that the attack surface grows alongside the number of devices connected to the network and the applications installed. Defending large attack surfaces is cumbersome and requires extensive maintenance. While IDS, Firewalls and other security appliances enable a high degree of protection and are key to enhance the overall security of a network, they may also contain vulnerabilities, hence, further increasing the attack surface~\cite{bellekens2014glop}. Deception-based defenses on the other hand enable a system to trick or mislead an attacker to perform (or not) an action. Two of the most widely accepted definition of deception applied to cyber-security are provided by Yuill~\cite{yuill2007defensive} and Almeshekah~and~Spafford~\cite{almeshekah2014planning}. In their manuscript, Almeshekah~and~Spafford discuss the addition of ``confusion" to the widely accepted definition provided by Yuill. 
In this paper we argue, that deception and confusion in cyber-security is best achieved, through human computer interaction. Static systems, such as described in literature~\cite{nawrocki2016survey}~\cite{han2018deception} often fail to meet the definitions aforementioned. We further argue that high interaction environments are key to deceptive defenses. Low interaction deceptive environments fail to mimic the behaviour of virtualised systems, furthermore, they fail to provide malicious user with a viable challenge and hence, limiting the usefulness of such systems~\cite{merien2018spatio}. Our platform, while providing a fully interactive system, further extends this by building upon gamification techniques and story telling (narrative), intrinsically, providing an immersive environment to the malicious user similar to video games. 

\subsection{Gamification}
Interactive defences, such as honeypots, are often based on real or virtualised systems logging interaction information and queries made by malicious users. While honeypots are effective at detecting and deflecting cyber-attacks or creating malware zoo, their deceptive action is often seen as a secondary purpose~\cite{8602913}. Furthermore, while enabling interaction honeypots are not designed to coerce the user into further interacting with it. To this end, we challenged the definition of high interaction honeypots, by combining high interaction environments with gamification techniques, coercing malicious users into participating in an unknown (to the malicious user) challenge. A generally accepted definition of gamification is provided by Deterding~\textit{et al.} as \textit{``The use of game design elements in non-game contexts"}~\cite{deterding2011game}. The gamification and design techniques used within the interactive deception platform are derived from~\cite{blohm2013gamification}. 

\paragraph{Game Mechanics} act as building blocks, for example, the scoring systems, badges and trophies enable the user to rank himself and foster the motivation of the user to perform better. In the platform, the scoring system is enabled by providing the user with the ability to compare his achievements against planted hacker achievements~\cite{blohm2013gamification}.
 
\paragraph{}Gamification and Game Mechanics must be tightly coupled with a game narrative in order to provide best results and ensure immersion. 

\subsection{Narrative}
In order to foster interaction with the deception platform , the platform builds on an attack narrative, additionally to the gamification techniques used. The narrative enables to construct a causal, logical and chronological chain of events, necessary to the high interaction capabilities of the platform. Barthes~\cite{barthes1979lecture} described events critical to the coherence of the story as ``cardinal functions". Cardinal functions play an important role within story telling as it enables branching into sub-stories, but also allows the malicious user to immerse himself within the challenge provided by the platform. In order to increase the deception aspect of the platform, the narrative follows the intrinsic motivation of a malicious user. Furthermore, the narrative must be plausible, as different network architecture may have different vulnerabilities and rewards, and hence, intrinsic motivations. Furthermore, narratives must cultivate the curiosity of the malicious user as the story unfolds, it is therefore important to align the narrative to the gamification aspect of the platform.
To this end, the platform provides a believable chain of events based on potential cyber-attacks and vulnerabilities induced within the systems as scenarios. The stories provided by the platform are linear, and constructed retrospectively, in order to avoid mismatches between the vulnerabilities, the gamification, the rewards and the narrative. 
\newline
The gamification and reward fully inform the narrative, guiding the malicious user to its next step. While the malicious user can perform an infinite number of action, he is unable to influence the outcome of the game without following the path laid out. 

\subsection{Manipulation}
A unique approach to deceive and mislead users is through manipulation. Manipulation can be defined as the action of controlling someone at your advantage. Manipulation can take different forms as explored by Buss~\cite{buss1992manipulation}. In his study, Buss reported 12 different manipulation techniques used in close relationships and interactional context. We identified 7 techniques to work with our deception platform, each leading the malicious user to perform an action at our advantage to either follow the story or to follow another path (i.e. target). 

\paragraph{Coercion} is the practice of making someone perform a forced action by using threat or force. Through data gathering, and behavioural analysis an operator is able to modify the narrative and gamification element relating to a specific malicious user, enabling to display coercive error message, forcing the hacker to follow any other pre-defined given path. 

\paragraph{Reciprocity Reward} are provided through a flexible narrative. As aforementioned, the platform can be used to monitor the actions of a malicious user. In order to evaluate the user's competencies, the platform requires manual vulnerabilities testing, hence, the gamification engine provides reciprocal rewards to users that are not using vulnerabilities scanners. Enabling them to continue further discovering key elements within the deceptive environment.

\paragraph{Debasement} is the practice of lowering values, to manipulate someone. While unable to work on the emotions of the malicious user, the narrative is able changes by analysing the attacks performed. Vulnerabilities can be loaded in a round robin fashion, in order to keep the malicious user entertained, and therefore exhibit less difficulties. Debasing the platform challenges ensures the continuation of the story and thus the immersion of the malicious user within the deceptive game. 

\paragraph{Charm} focuses on complimenting someone in order to perform an action. As the platform and the narrative require a logical and believable construct, charming the malicious user is only done through subtle clues, such as comments on the ``ability" of a malicious user in code running on the client-side. Those charming ``comments" help the user follow the pre-defined path, while believing to follow this path through is own free will. 

\paragraph{Pleasure Induction} is the ability of demonstrating that an unwanted action can be pleasurable for the malicious user. This is provided by the ability to debase specific vulnerabilities enabling the malicious user to continue is progression through the deceptive platform and earn rewards. 

\paragraph{Social Comparison} is the act of comparing the malicious user against other hackers or groups. Due to the nature of the platform, obvious comparisons such as leaderboard would annihilate the immersive element of the platform. To this end, comments on client-side and defacing pages are hidden. Website defacement is an attack that changes the appearance of a website, often accompanied by the name of the defacers or the group of defacers. This subtle clue, enables to create a social comparison between the malicious user and other groups or hackers. 

\paragraph{Monetary Rewards} are core of some scenarios to deceive the hacker. The narrative builds upon potential credit cards, or valuable information available at the end of the quest, hence leading the attacker to follow the narrative to obtain a potential monetary reward. 

\subsection{Gratification} 
Luring a malicious user into deception requires to foster its curiosity, in this manuscript, we use gratification  to trigger a pleasurable emotional reaction within the malicious user, in order to lure him to perform key deceptive actions. By providing the malicious user, with credit card information, rewards, or key information, we lead the action of the user towards a specific path or target predefined within the platform. The gratification inherently focuses on increasing the deceptive action of the platform.  

\section{Platform Architecture}
\label{sec:PlatformArchitecture}
This section provides an overview of the deceptive platform architecture. We developed a fully extendable gamification architecture allowing researchers to deploy virtualised hosts on both local networks and the internet. Each virtualised hosts contains a set of specific vulnerability~(i.e. web application, software, buffer overflow, etc). Each host is connected to a game engine, an interaction engine and a scoring engine.
     
\begin{figure}[h]
    \centering
        \includegraphics[height=0.5\textheight, keepaspectratio=true]{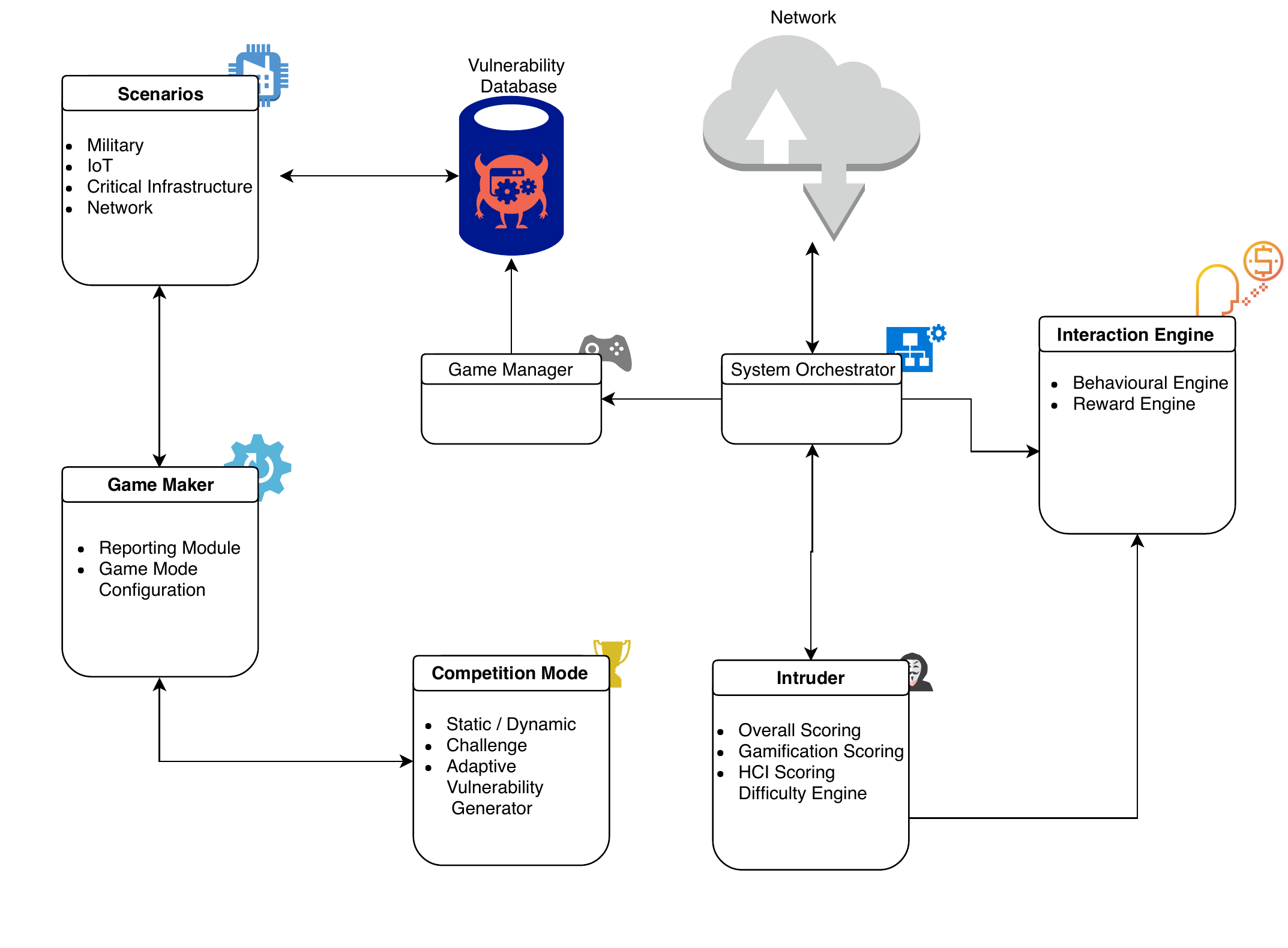}
        \caption{Logical Architecture of the Deception Framework}
      \label{LADF}
\end{figure}
\textbf{}
Figure~\ref{LADF} depicts the logical architecture of the platform. When a hacker connects to one or more virtual hosts, he is unable to differentiate it from a real-world computer (i.e. running a windows operating system), this is achieved by using port-scanning deception enabling to camouflage the signature of the operating system. 

All interactions with the host(s)~including time, behavioural data~(i.e. Keystroke dynamics, activity tracking, etc.), and engagement are further recorded and processed by the game and the scoring engines. This allows the hacker to be served with polymorphic vulnerabilities which, in turn, can increase or decrease their difficulties over time, keeping the hacker engaged with the platform. Furthermore, Figure~\ref{LADF} shows that the interaction information gathered through the host(s) is fed to the scoring engine, which provides the hacker with rewards based on pre-defined scenarios. Using a threshold measures, the hacker's interest is further analysed. If his interest scores below threshold, subtle clues are provided to the hacker. The clues are inbuilt in each scenario. The clues vary from wireshark captures, to misleading network scans and vulnerability scans. The clues enable the hacker to seamlessly continue his malicious activity on the the network by following a pre-defined path, without suspecting interacting with a virtual environment. The path leads to data being gathered on the attacks, techniques and tools used by hackers to solve each challenges thrown at him. All the gathered information are further analyzed using a circular methodology, enabling the operators to enhance the game engine and the variability of the difficulties. 
 
\begin{figure}[h]
  \centering
        \includegraphics[height=0.35\textheight, keepaspectratio=true]{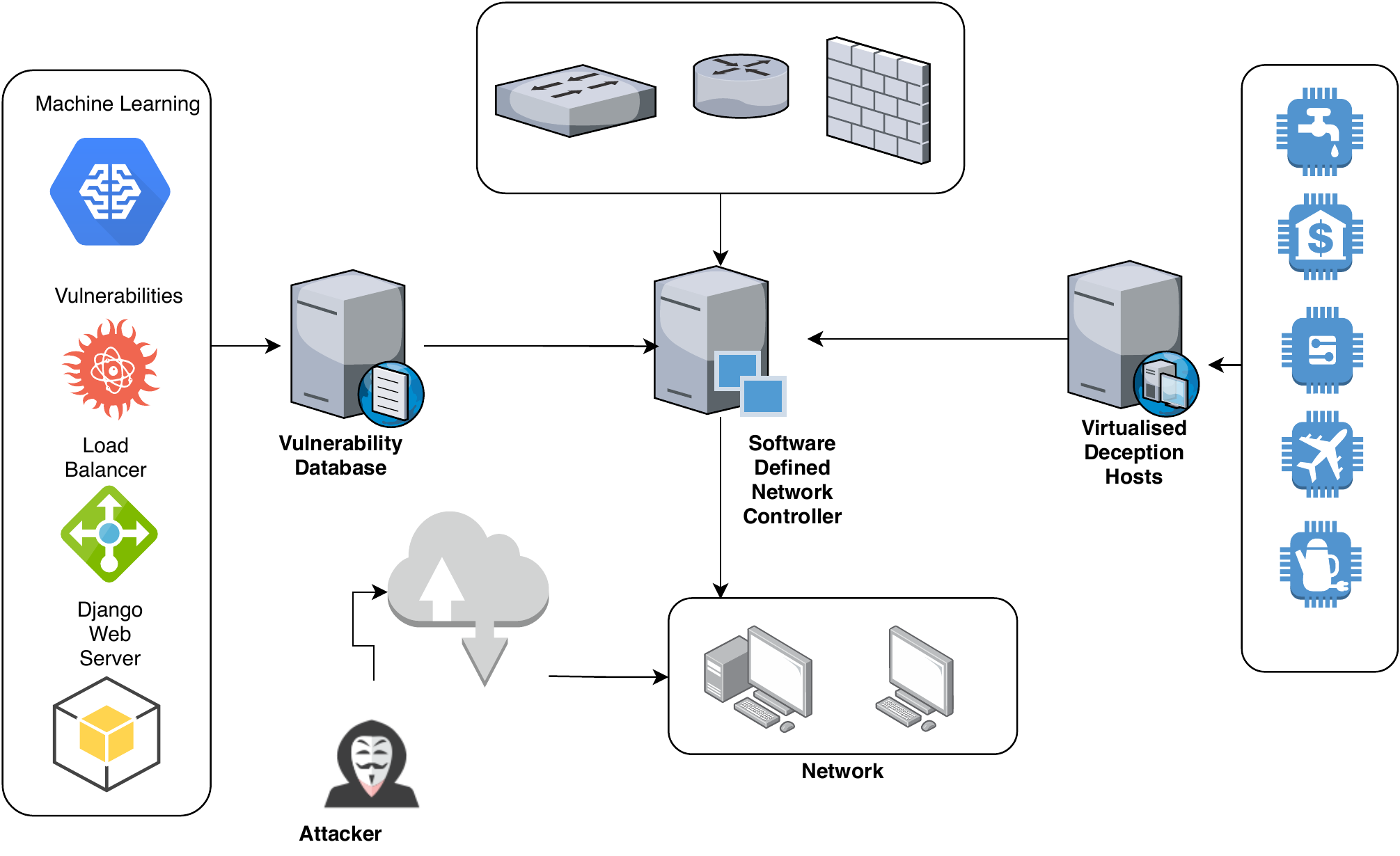}
        \caption{Network the Deception Framework}
        \label{ADF}
\end{figure}

Figure~\ref{ADF} provides an overview of the deception framework network architecture. The network can be extended through the Software Defined Network~(SDN) controller, enabling an operator to create virtual networks on demand. Information are pulled from the vulnerability database to populate the virtualised hosts running on the virtualised deception host server. Based on the scenario being launched, different network architectures, virtual hosts and vulnerabilities are loaded and spawned on the network. Hacker interactions are made through a load balancer and a Django web server for all web vulnerabilities. 

\section{Scenarios}
\label{sec:Scenarios}

\subsection{Scenario 1: Shopping Website} 
77\% of web applications have been reported to include a vulnerable JavaScript library enabling malicious users to take advantage of the website and potentially use this vulnerability to launch further attacks. The scenario was evaluated on its own, to identify the state of the narrative, the value of the kernels within the narrative, and the deception, manipulation and gamification aspects.

\begin{figure}[h]
  \centering
        \includegraphics[height=0.42\textheight, keepaspectratio=true]{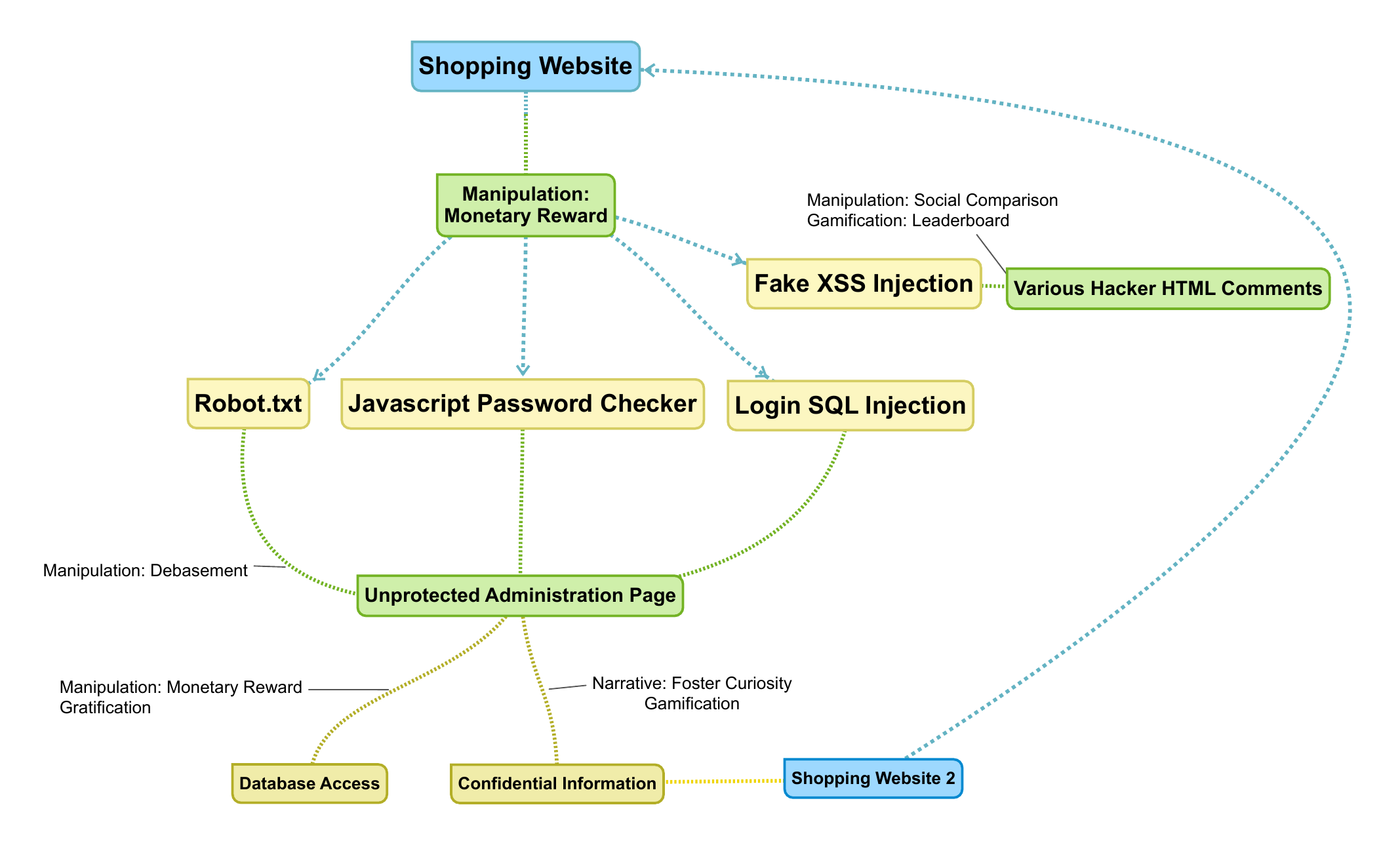}
        \caption{Web Deception Narrative}
        \label{WEBDN}
\end{figure}

\paragraph{}Figure~\ref{WEBDN} provides an overview of the deception narrative. The shopping website offers potential monetary rewards if hacked through credit cards stored in the database. To this this end, 4 apparent vulnerabilities were created. I) A ``robot.txt" wile containing the path to an unprotected administration page. II) A ``JavaScript password checker" library included within the ``index.html" page enabling the hacker to bypass the login and password requirements. III) A ``Login SQL Injection" enabling the hacker to use an SQL injection against the password field to gain access to the administration page. IV) A persistent cross-site scripting~(XSS) injection vulnerability.  

\paragraph{}Through subtle clues, within the code, comments, etc\ldots, the attacker is lead to believe that the website is vulnerable. His interested is further piqued through the manipulation techniques as shown on the edges of Figure~\ref{WEBDN}. By following the XSS injection vulnerability, the malicious user will quickly discover planted comments from other imaginary hackers, acting as a leaderboard and creating a social comparison between the abilities of the imaginary hackers and the malicious user.  Furthermore, by following the first 3 vulnerabilities, the hacker will access an unprotected administration page, which, in turn, will give him access to a ``database.db" file containing fabricated credit card numbers and associated CSV numbers. Another file provides information about another shopping website, which simulate similar vulnerabilities with a different design, misleading the hacker to follow another path.

\subsection{Scenario 2: Vulnerable FTP} 
The second scenario is built around a vulnerable version of an FTP application. The FTP contains default credentials enabling a malicious user to read and upload new data on the FTP. The premise of this scenario is to reward the hacker upfront to provide direct gratification.  

\begin{figure}[h]
  \centering
        \includegraphics[height=0.25\textheight, keepaspectratio=true]{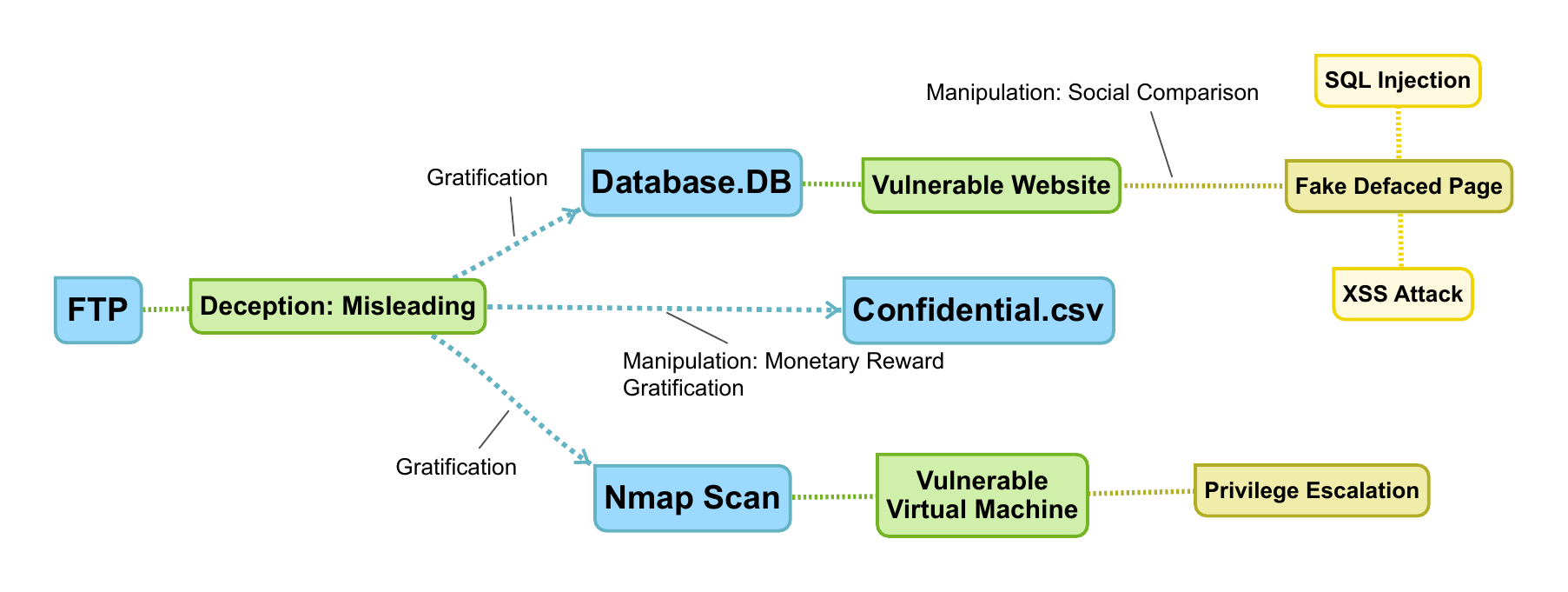}
        \caption{FTP Deception Narrative}
        \label{FTPDN}
\end{figure}

\paragraph{}Figure~\ref{FTPDN} illustrates the different path the malicious user is able to take. Providing upfront gratification, with clear planted rewards enable to build the momentum of the narrative. The first file is a database file containing various tables and information. The file contains 2GB of data, requiring the hacker to spend time analysing and searching for potential information. Within the file, numerous references to a vulnerable WordPress website are made. The link to a defaced page is also provided. The WordPress website is itself vulnerable to SQL injection and XSS attacks and aims at comparing the malicious user skills with the skills of an imaginary planted hacker. The second vulnerability provides the malicious user with a large Nmap scan of numerous machine. One is identified as vulnerable. Upon exploiting the vulnerability the malicious user is able to perform a privilege escalation through a buffer overflow attack (All buffer overflow protections are disabled).

\subsection{Scenario 3: IoT Network Combination}
This scenario builds upon Scenario 1 and 2, by integrating them into a vulnerable IoT Network running MQTT nodes. The aggregation of scenarios enable to construct a plausible interactive deception environment, leading the malicious user towards predefined targets and exhaust his skills against known vulnerable nodes allowing operators to segregate the network, essentially, exploiting the malicious user for the benefit of data gathering. 

\begin{figure}[h]
  \centering
        \includegraphics[height=0.45\textheight, keepaspectratio=true]{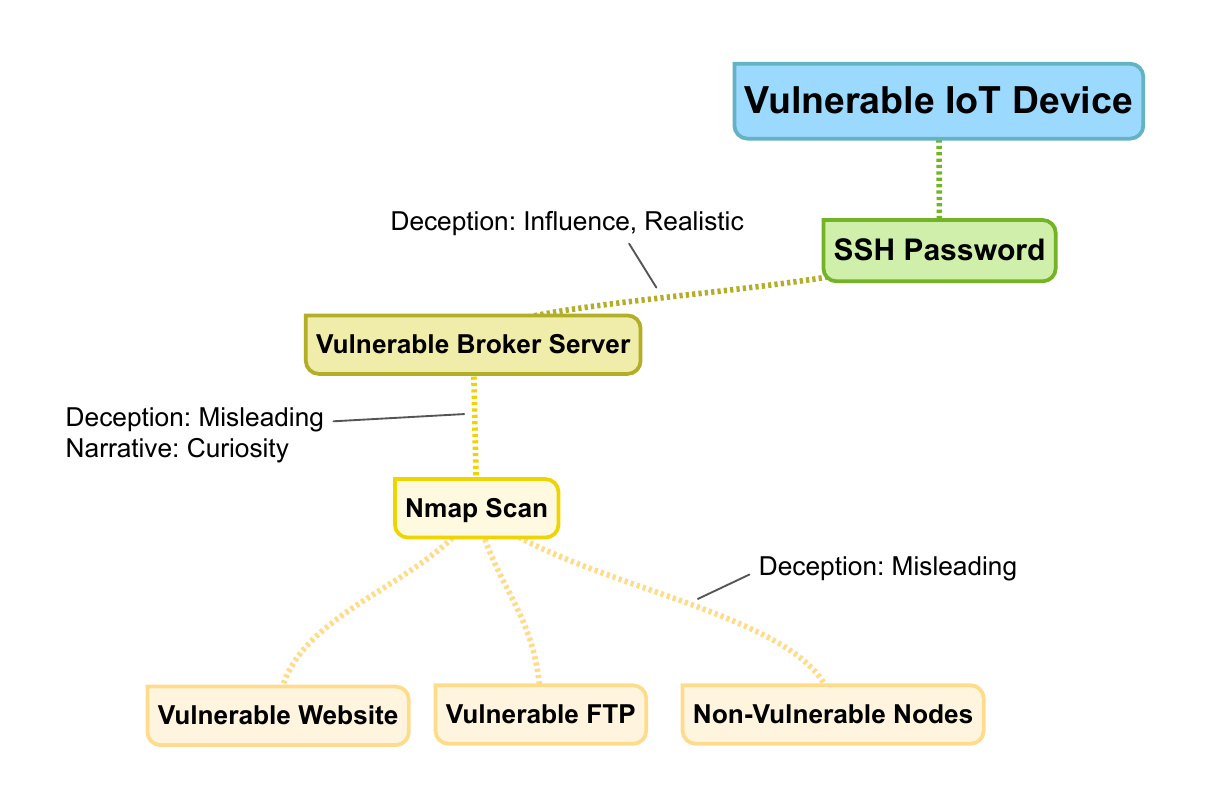}
        \caption{IoT Network Deception Narrative}
        \label{IoTDN}
\end{figure}

\paragraph{}Figure~\ref{IoTDN} provides an overview of the vulnerable IoT network. One front facing vulnerable IoT device running MQTT uses weak SSH credentials. The device itself contains a file with un-encrypted SSH credentials for a vulnerable ``broker server". The root of the broker contains an NMAP Scan which has for objective to lead the malicious user towards the vulnerable devices analysed. The scan contains information about 7 nodes in total. One vulnerable webserver, one vulnerable FTP and 5 IoT nodes with weak credentials but advertised within the NMAP scan as no port open and secured. Using Nmap as an ``authority figure" to mislead the user into selecting another proposed target. 

\section{Evaluation}
\label{sec:Evaluation}

The platform was built against distinctive scenarios evaluating the narrative, the gamification and the manipulation aspects of the platform. Leading the attacker on a pre-defined path for deception.

\paragraph{} All deception environments have been launched front facing the internet using four different locations (Atlanta, Lyon, London, and Tokyo) for 2 weeks. All interactions with the environment were recorded and the path of the hackers where analysed for each environment. For consistency, all environments were reset at midnight (local time). The path taken by malicious users accessing different systems are recorded using their IP addresses. Furthermore, all malicious users following a path on a predefined scenario have been aggregated.

\begin{figure}[h]
  \centering
        \includegraphics[height=0.45\textheight, keepaspectratio=true]{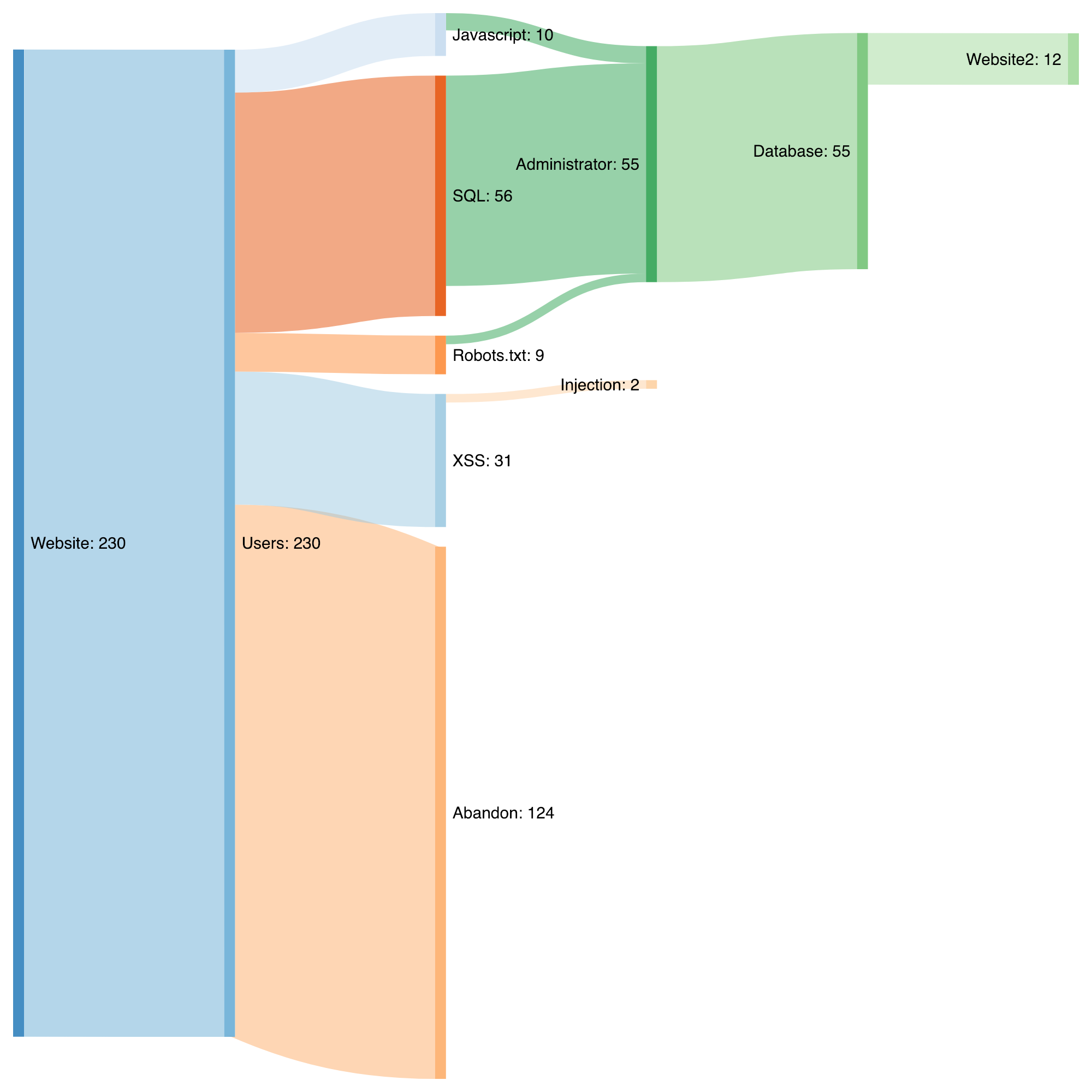}
        \caption{Website Flow Narrative}
        \label{WFN}
\end{figure}

Figure~\ref{WFN} shows the path followed by malicious users while interacting with the interactive deception web Scenario. A total of 230 users accessed the platform over a period of two weeks. 53.9\% of the users did not succeed into accessing the administrative web page, note that in this case no distinctions are made between potential legitimate users and unsuccessful malicious users and bots. 13.4\% of malicious users tried XSS injection, but only 0.86\% resulted in a successful persistent XSS injection. 3.9\% of the users accessed the ``robots.txt" file and only 0.86\% further accessed the administrator webpage. A surprising result, as this clue was created for debasement purpose (i.e. enabling easy access to the second stage). 24.3\% of malicious users performed  an SQL attack against the website, however only 23.9\% succeeded into accessing the administrator page. 4.34\% accessed the JavaScript library, however, only 1.73\% of users made it to the administrator page. All users having access to the administrator webpage accessed the database file, however, 21\% of the successful users went on to attacking the second deceptive website.

\paragraph{} While the drop-out rate is high, 46\% of users carried out at least one attack, with a total of 27.9\% carrying out a successful attack. The average time spend by user 29.6 minutes. The flows presented in Figure~\ref{WFN} indicate a successful narrative, as successful users tried to proceed with an attack on a pre-defined path, the drop-outs at various stages are indicated by a lack of motivation (i.e. potential gratification) or a lack of skills.

\begin{figure}[h]
  \centering
\includegraphics[height=0.45\textheight, keepaspectratio=true]{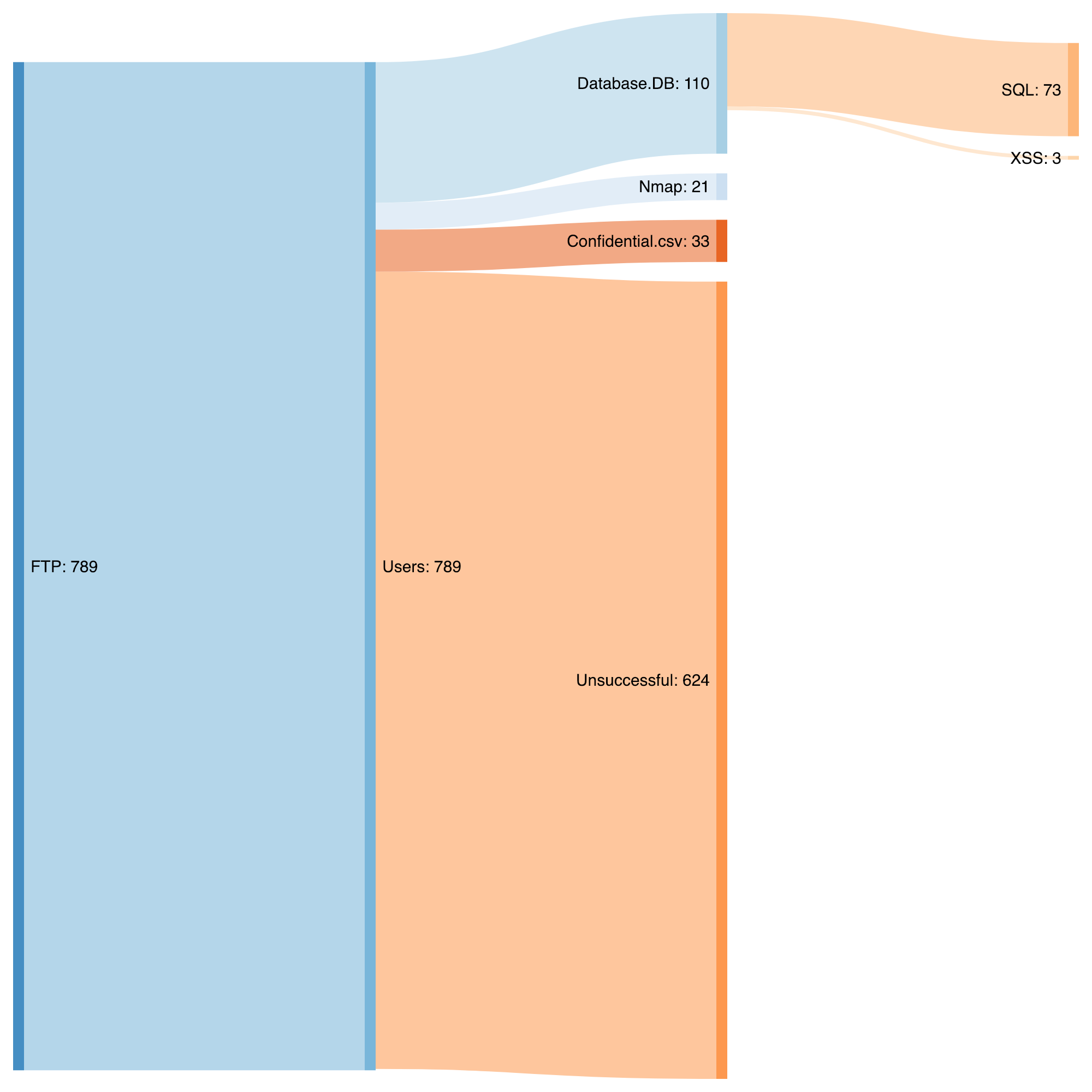}
        \caption{FTP Flow Narrative}
        \label{FFN}
\end{figure}

Figure~\ref{FFN} illustrates the path taken by malicious users during their interaction with the interactive deception FTP Scenario. A total of 789 IP addresses were recorded trying to access the FTP over a two week period. 20.7\% malicious users successfully logged on the FTP server. 67\% of successful users accessed the ``Database.DB" file, further leading them to access a planted vulnerable website. 66.36\% carried out an SQL attack against the website, while 2.72\% carried out an XSS attack. 12.8\% of the successful users (at the first stage) accessed the NMAP scan, but none of them carried out any attacks against the vulnerable virtual machine. However 20.12\% of the malicious users (at the first stage) accessed the ``confidential.csv" containing, fake names, credit card numbers and CSV numbers. 

\paragraph{}The primary assumption for the high-drop off at the first stage is due to the number of bots that scanned the FTP and abandoned. The low number reported to access the ``confidential.csv" file is believed to be due to the too obvious name of the file, hence leading malicious users onto pursuing the ``database.DB" file and executing SQL and XSS attacks against the website. Malicious users spend an average time of 9.36 minutes on the platform.

\begin{figure}[h]
  \centering
\includegraphics[height=0.45\textheight, keepaspectratio=true]{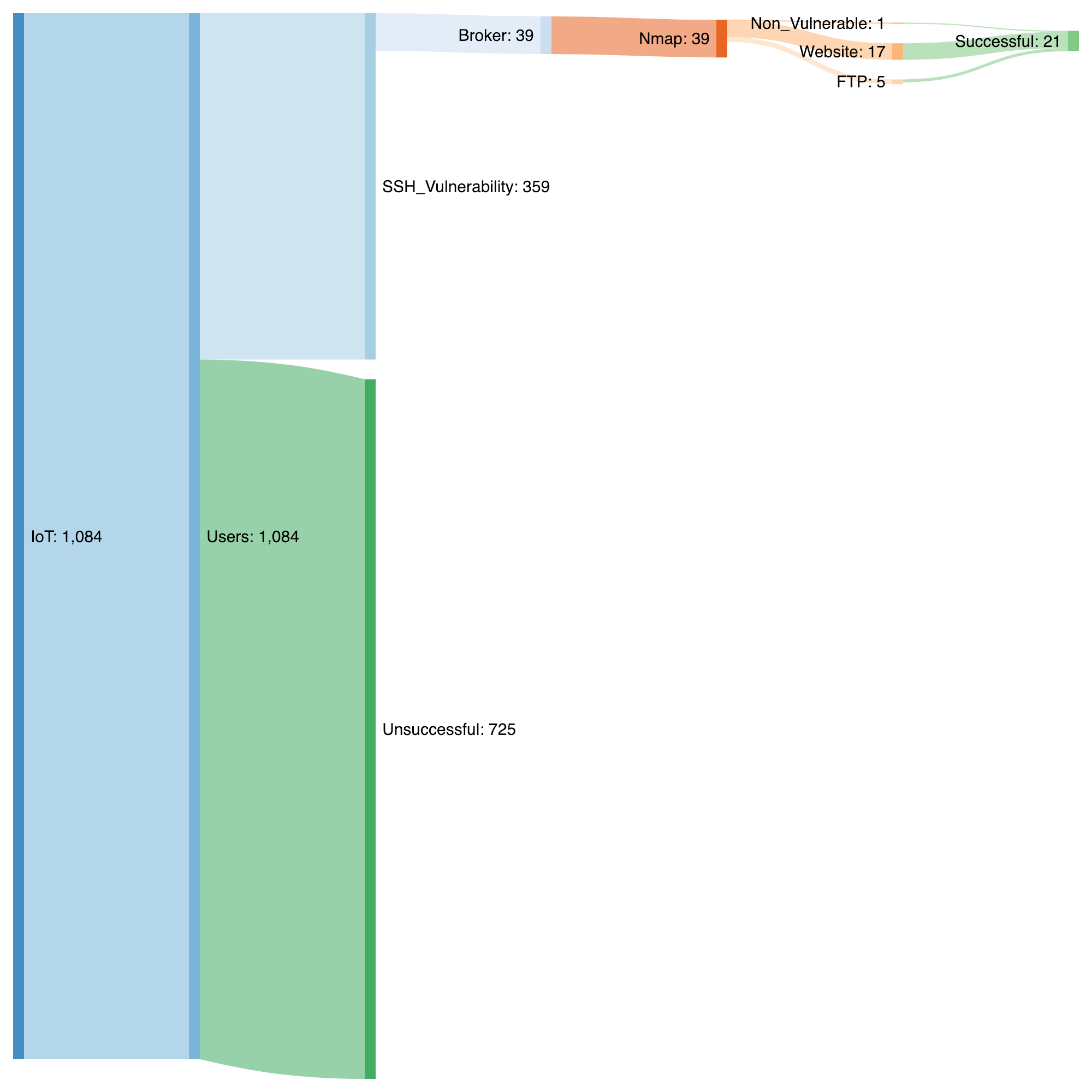}
        \caption{IoT Flow Narrative}
        \label{IOTS}
\end{figure}

Figure~\ref{IOTS} shows the path taken by malicious users when interacting with the combined environment, through an IoT vulnerable device sending MQTT requests to a vulnerable broker. Out of the 1,084 individual IP recorded 66.88\% of requests made to the IoT platform were unsuccessful. 33.11\% of attacks succeeded into accessing the IoT virtual sensor by identifying the weak credential uses, however, only 10.8\% of the successful users where able to connect to the broker. 100\% of the users accessing the broker (stage 3) accessed the Nmap file stored at the root. 2.59\% of the successful malicious user accessing the nmap file, attacked the nodes stated as non-vulnerable in the Nmap scan, and successfully accessed the 5 nodes. 56.4 \% of the users reacted to the misleading information and attacked either the website or the FTP. All malicious users that proceeded to stage 4 performed successful attacks against the deceptive targets. 43.5\% also dropped off at stage 3 after having accessed the Nmap file. It is expected that the drop off, might be due to a lack of reward at this stage,  clarity in reading the Nmap file provided or a \textbf{}lack of time. The malicious users spend an average time of 41.3 minutes on the deceptive scenario demonstrating a successful, believable narrative, fostering curiosity.  

\section{Related Works}
\label{sec:RelatedWorks}
McQueen~\textit{et al.}~\cite{mcqueen2009deception} introduced deception defenses in control systems. They defined security through seven abstract concepts, and further defined deception solutions for each abstract security concept. The concepts and associated deception are provided in the form of a broad taxonomy. Heckman~\textit{et al.}~\cite{heckman2013active} review the January 2012 real-time red/blue team wargame experiment by MITRE, which included deception at its core, in order to fool and deceive red teams. The main focus of the deception system in place was to simulate a command and control system, which in turn would provide false information to the read team through a fake interface. The system presented by the authors was specifically designed with a wargame in mind, and while proving its efficacy, the system lacks flexibility in order to be deployed in a broader context (i.e. IoT, SME Networks, SCADA Systems, Web, etc\ldots). In \cite{stech2011scientometrics}~Stech~\textit{et al.} provide a scientometric analysis of the concepts of deception detection in cyber space. The authors demonstrated that the social, behavioral and cognitive aspects of deception where often discarded from the denial and deception tactics used. They also highlighted a lack of terminology to describe deception and classified deception as an emerging field within cyber-security.

\section{Discussion and Limitations}
\label{sec:KeyTakeAways}
This section introduces the key takeaways of the interactive deception platform as well as the limitations of the platform in its current state. 

\subsection{Key Takeaways}
\begin{itemize}
\item \textbf{Increased Interaction:} The platform enables malicious users to interact with numerous believable components and focuses on increasing the interaction time between the malicious user and the components. 
\item \textbf{Chronological Events:} The platform builds on the concepts of gamification and narrative, hence, unfolds in a time line fashion.  
\item \textbf{Deflection:} The use of gamification, manipulation, gratification and narrative have demonstrated to be successful in order to deflect attacks from key network components to secondary nodes, or towards the dedicated virtual environment.  
\item \textbf{Platform Flexibility:} The platform is modular and hence enables to fit, and build on numerous scenarios. 
\end{itemize}

\subsection{Limitations}
\begin{itemize}
\item \textbf{Scenario Complexity:} Complex environments may require complex scenarios, which may clutter the environments. Furthermore, building believable scenarios and a plausible narrative is time consuming. 
\item \textbf{Distinguishing between legitimate and malicious users:} In its current state the platform is unable to distinguish between users, hence limiting the interaction of the environment with legitimate users. 
\end{itemize}

\section{Conclusion}
\label{sec:Conclusion}
In this paper a successful interactive deception platform based on gamification, narrative, manipulation  and gratification techniques is presented. The platform is highly modular and enable users to develop and deploy scenarios to mislead and deceive malicious users on a network, to safeguard key elements of the network, buy time for security operators to isolate the deceptive nodes and malicious user, or to collect and study the behaviour of malicious users. We have also presented a comprehensive model to increase the interaction with deceptive platform. The techniques can easily be transferred across to existing platform. Furthermore, the scenarios presented are based on a plausible narrative encouraging and coercing the user into exploring the environment. In this paper 3 scenarios were presented, demonstrating the efficiency of the the narrative, manipulative and gratificative components. It was demonstrated that while 2103 individual connection to the platform were recorded over the three scenarios and a high initial drop-out rate, over 75\% of users engaging with the platform carry at least one successful attack.

%
 \bibliographystyle{splncs04}
 \bibliography{bibliography}

\end{document}